\documentclass[twocolumn,amsmath,aps]{revtex4}

\usepackage{amssymb,mathrsfs,bm,color,times}
\usepackage{graphicx,color}
\usepackage{bm}
\usepackage[hypertex]{hyperref}
\usepackage{amsmath}

\newcommand{\be}{\begin{equation}}
\newcommand{\ee}{\end{equation}}
\newcommand{\bea}{\begin{eqnarray}}
\newcommand{\eea}{\end{eqnarray}}
\newcommand{\bsube}{\begin{subequations}}
\newcommand{\esube}{\end{subequations}}

\newcommand{\Eq}[1]{Eq.\,(\ref{#1})}
\newcommand{\Eqs}[1]{Eqs.\,(\ref{#1})}

\newcommand{\la}{\langle}
\newcommand{\ra}{\rangle}

\newcommand{\beq}{\begin{equation}}
\newcommand{\eeq}{\end{equation}}
\newcommand{\beqn}{\begin{eqnarray}}
\newcommand{\eeqn}{\end{eqnarray}}

\newcommand{\nl}{\nonumber \\}

\newcommand{\bsub}{\begin{subequations}}
\newcommand{\esub}{\end{subequations}}

\begin{document}

\title{Quantum-coherence-free precision metrology
by means of difference-signal amplification}


\author{Jialin Li}
\affiliation{Center for Joint Quantum Studies and Department of Physics,
School of Science, \\ Tianjin University, Tianjin 300072, China}
\author{Yazhi Niu}
\affiliation{Center for Joint Quantum Studies and Department of Physics,
School of Science, \\ Tianjin University, Tianjin 300072, China}
\author{Xinyi Wang}
\affiliation{Center for Joint Quantum Studies and Department of Physics,
School of Science, \\ Tianjin University, Tianjin 300072, China}
\author{Lupei Qin}
\email{qinlupei@tju.edu.cn}
\affiliation{Center for Joint Quantum Studies and Department of Physics,
School of Science, \\ Tianjin University, Tianjin 300072, China}
\author{Xin-Qi Li}
\email{xinqi.li@tju.edu.cn}
\affiliation{Center for Joint Quantum Studies and Department of Physics,
School of Science, \\ Tianjin University, Tianjin 300072, China}

\date{\today}

\begin{abstract}
{\flushleft
The novel} weak-value-amplification (WVA) scheme of precision metrology
is deeply rooted in the quantum nature of destructive interference
between the pre- and post-selection states.
And, an alternative version, termed as joint WVA (JWVA),
which employs the difference-signal from the post-selection
accepted and rejected results,
has been found possible to achieve even better sensitivity
(two orders of magnitude higher)
under some technical limitations (e.g. misalignment errors).
In this work, after erasing the quantum coherence,
we analyze the difference-signal amplification (DSA) technique,
which serves as a classical counterpart of the JWVA,
and show that similar amplification effect can be achieved.
We obtain a simple expression for the amplified signal,
carry out characterization of precision,
and point out the optimal working regime.
We also discuss how to implement the post-selection
of a classical mixed state.
The proposed classical DSA technique holds
similar technical advantages of the JWVA
and may find interesting applications in practice.
\end{abstract}


\maketitle

{\flushleft
Applying} the concept of quantum weak values (WVs)
proposed by Aharonov, Albert and Vaidman (AAV) \cite{AAV88,AV90},
a novel scheme of precision metrology termed as weak-value amplification (WVA)
has caused great interest over the past decade and a half
\cite{Kwi08,How09a,How09b,How10a,How10b,Guo13,Sim10,Ste11,Nish12,Ked12,
Jor14,Li20,Li22a,Bru15,Bru16,How17,Lun17,ZLJ20,Jor13,Sim15,Jor21}.
The WVA technique allows probe sensitivity beyond the detector's resolution
and can outperform conventional measurement in the presence
of detector saturation and technical imperfections \cite{Lun17,ZLJ20}.
The WVA technique involves an essential procedure termed as post-selection,
which discards a large portion of output data.
Physically speaking, the WVA
is rooted in the quantum nature of
interference effect
between the pre- and post-selected (PPS) states.
In the singular amplification regime,
this novel quantum effect allows the WVA measurement
to put almost all of the Fisher information
about the parameter under estimation
into the small portion of the remained data \cite{Ked12,Jor14,Li20,Li22a},
which leads thus to some important technical advantages.

However, aside from the singular amplification regime,
viewing that the discarding data by post-selection
encode considerable information,
a different strategy of amplification was proposed
by considering to use all the post-selection accepted (PSA)
and rejected (PSR) data \cite{Bru13,ABWV16,ABWV17a,ABWV17b,Zeng16,Zeng19,Li22}.
This proposal was referred to as joint-weak-measurement or joint WVA (JWVA) scheme.
Importantly, it was argued that the JWVA scheme permits the removal
of systematic error, background noise, and fluctuations in alignments
of the experimental setup \cite{ABWV17a,ABWV17b,Zeng16}.
In Ref.\ \cite{ABWV17b}, it was demonstrated that the JWVA
offers on average a twice better signal-to-noise ratio (SNR)
than WVA for measurements of linear velocities;
while in Ref.\ \cite{Zeng19}, the JWVA was estimated
having a sensitivity two orders of magnitude higher than the WVA,
under some technical imperfections (e.g. misalignment errors).

Being different from the standard WVA, in the JWVA scheme,
the intensities of the PSA and PSR signals are set almost equal
and the difference between them reveals anomalous amplification \cite{Bru13,ABWV16}.
In present work, along the same line of subtracting the PSA and PSR signals,
but erasing the quantum coherence in the PPS states,
we consider the classical counterpart of the JWVA
and name it difference-signal amplification (DSA) scheme.
This is motivated by noting that the amplification principle of the JWVA
is largely based on a statistical trick, but not on the quantum interference effect.
We thus conjecture the possibility of developing a quantum-coherence-free DSA technique,
which holds similar advantages of the JWVA in the presence of technical imperfections
such as systematic errors and misalignment limitations.

In this work, as a theoretical model, we employ the Stern-Gerlach setup
but erase the quantum coherence of the electron spin.
After coarse graining treatment \cite{Li16},
it coincides with the classical coin-toss model analyzed in Ref.\ \cite{FC14b},
where it was argued that the coin-toss model
can generate also the effect of anomalous WV,
if introducing proper external noise (disturbance).
Then, it was concluded that quantum interference
is not the unique reason for the AAV's anomalous WV;
in contrast the anomalous WV is largely owing to a statistical procedure \cite{FC14b}.
In Ref.\ \cite{Li16}, it was clarified that
in classical system (without quantum coherence)
it is impossible to generate the AAV's anomalous WV by post-selection.
Therefore, we may remind to distinguish the DSA under present study
from the ``disturbing" noise treatment in Ref.\ \cite{FC14b}.  \\
\\
{\flushleft\bf Formulation of the DSA.}$~~$
Let us start with the standard Stern-Gerlach setup,
which describes in general a quantum two-state system
coupled to a meter for weak measurement \cite{AAV88,AV90}.
The interaction between the system (electron spin)
and the meter (electron's transverse spatial degrees of freedom)
can be described by $H'=\kappa P A$,
with $P$ the momentum operator and $A=\sigma_z$ the Pauli operator for the spin.
In quantum case, the spin of electron
is initially prepared in a quantum superposition
\bea
|i\ra= \alpha |\uparrow\ra  + \beta |\downarrow\ra  \,,
\eea
with $|\uparrow\ra$ and $|\downarrow\ra$ the spin-up and spin-down states.
The electron's transverse spatial wavepacket is assumed to be a Gaussian
\bea
\Phi_0(x)=\frac{1}{(2\pi \sigma^2)^{1/4}}
\exp\left[-\frac{x^2}{4\sigma^2}\right],
\eea
with $\sigma$ the width of the wavepacket.
After passing through the area of the inhomogeneous magnetic field
in the Stern-Gerlach setup,
the electron's spatial wavepacket would experience two possible shifts,
becoming as
\bea\label{Phijx}
\Phi_{\uparrow,\downarrow}(x)=\frac{1}{(2\pi \sigma^2)^{1/4}}
      \exp\left[-\frac{(x-\bar{x}_{\uparrow,\downarrow})^2}{4\sigma^2} \right] \,,
\eea
where $\bar{x}_{\uparrow,\downarrow}=\pm d$ are the respective Gaussian centers
shifted by the coupling interaction $e^{-id P A}$.
The parameter $d=\int_0^{\tau} dt\, \kappa=\kappa \tau$,
with $\tau$ the interaction time, is what we want to estimate
through measuring the spatial wavepacket.

In quantum case, if we perform a post-selection to the spin state with,
for example, $|f\ra= a |\uparrow\ra + b |\downarrow\ra$,
under the limit of weak measurement strength $g=(d/2\sigma)^2<<1$,
the electron's spatial wavepacket would experience a shift
from $\Phi_0(x)$ to $\Phi_0(x-A_w d)$, with
\bea
A_w=\frac{\la f|A|i\ra}{\la f|i\ra} \,.
\eea
This is the well-known AAV's WV.
One can check that the WV $A_w$ can considerably exceed
the range of eigenvalue spectrum of the physical quantity $A$.
That is, proper post-selection can cause anomalous WV,
while the underlying reason is the {\it quantum interference}
between the PPS states $|i\ra$ and $|f\ra$.

To see this more clearly, let us reexpress the WV as
\bea
A_w &=& \frac{\langle f|A| i\rangle}
{\langle f| i\rangle}
= \frac{\langle f|A|i\rangle \langle i|f\rangle}
{\langle f|i\rangle \langle i|f\rangle}  \nl
&=& \frac{{\rm Tr}[\rho_f A\rho_i]}{{\rm Tr}[\rho_f\rho_i]}
= \frac{M_1}{M_2}  \,.
\eea
Here we have introduced the density matrices
$\rho_i=|i\ra\la i|$ and $\rho_f=|f\ra\la f|$,
for latter convenience of switching to consider classical states
(i.e. a statistical mixture of the spin-up and spin-down states).
Simple calculation yields $M_1=\alpha^2 a^2-\beta^2 b^2$
and $M_2=(\alpha a+\beta b)^2$.
Here we assume that the superposition coefficients
$(\alpha, \beta)$ and $(a,b)$ are real.
When $\alpha a+\beta b\rightarrow 0$, we find singular weak values.
Actually, this condition corresponds to
the {\it destructive interference} between the PPS states $|i\ra$ and $|f\ra$.
This is indeed a quantum effect, since the classical counterpart of $M_2$,
say, $M_2= \alpha^2 a^2+\beta^2 b^2$, can never be zero under any PPS choices.
Since the WVA technique is based on the amplification effect of the anomalous WV,
we know that in classical systems it is impossible to develop this same technique.
However, as to be shown in the following,
it is possible to develop a {\it quantum-coherence-free} amplification technique,
using both the PSA and PSR signals.

Realization of a classical state corresponds to erasing quantum coherence
from a quantum pure state \cite{Vaid17},
which changes the quantum superposition to a statistical mixture.
Then, let us consider the initial state of the electron spin
as a statistical mixture of the spin-up and spin-down states given by
\bea
\rho_i=\alpha^2 |\uparrow\ra \la\uparrow| + \beta^2 |\downarrow\ra \la\downarrow| \,.
\eea
The total state of the system-plus-meter before coupling interaction
is described as $\rho_T=\rho_i \otimes P_0(x)$.
Here $P_0(x)=|\Phi_0(x)|^2$, in classical case which corresponds to
the transverse spatial distribution of the particle beam
owing to stochastic emissions of the particles.
After coupling interaction, the $x$-measurement on the meter state
would change the electron's spin state as
\bea
\widetilde{\rho}_x=\alpha^2 P_{\uparrow}(x) |\uparrow\ra \la\uparrow|
+ \beta^2 P_{\downarrow}(x) |\downarrow\ra \la\downarrow|  \,,
\eea
where $P_{\uparrow,\downarrow}(x)=|\Phi_{\uparrow,\downarrow}(x)|^2$
are the spatial distributions shifted from $P_0(x)$.
Then, consider a post-selection for the $x$-measurement output data
with also a statistical mixed state, say,
$\rho_f=a^2 |\uparrow\ra \la\uparrow| + b^2 |\downarrow\ra \la\downarrow|$.
(How to realize this type of post-selection
is remained in the final overall discussion of this article.)
Theoretically, the distribution function of the PSA results is
\bea\label{P1}
&& \widetilde{P}_1(x)=\widetilde{P}(x;f)={\rm Tr}(\rho_f \widetilde{\rho}_x) \nl
&& ~~ = \alpha^2 a^2 P_{\uparrow}(x) + \beta^2 b^2 P_{\downarrow}(x) \,.
\eea
Accordingly, the distribution of the PSR results is obtained as
\bea\label{P2}
\widetilde{P}_2(x) &=& \widetilde{P}(x;\bar{f})
={\rm Tr}(\rho_{\bar{f}} \widetilde{\rho}_x) = P(x)-P_1(x)  \nl
&=& \bar{a}^2\alpha^2 P_{\uparrow}(x) + \bar{b}^2\beta^2 P_{\downarrow}(x) \,.
\eea
Here we have used $\rho_{\bar{f}}=1-\rho_f$
and introduced $\bar{a}^2=1-a^2$ and $~$ $\bar{b}^2=1-b^2$.
Further, the normalized distribution functions read as
$P_1(x)=\widetilde{P}_1(x)/p_f$ and $P_2(x)=\widetilde{P}_2(x)/p_{\bar{f}}$,
while $p_f=\alpha^2 a^2+\beta^2 b^2$ and
$p_{\bar{f}}=\alpha^2 \bar{a}^2+\beta^2 \bar{b}^2$
are the respective PSA and PSR probabilities.
Then, the expectation values of the PSA and PSA results
can be simply calculated as
\bea\label{xf-xfbar}
\la x\ra_f &=& \int dx x P_1(x) = (\alpha^2 a^2-\beta^2 b^2)d/p_f   \,,  \nl
\la x\ra_{\bar{f}} &=& \int dx x P_2(x)
= (\alpha^2 \bar{a}^2-\beta^2 \bar{b}^2)d/p_{\bar{f}}  \,.
\eea
Parameterizing the post-selection by introducing
$a^2=\cos^2\frac{\theta}{2}$ and $b^2 =\sin^2\frac{\theta}{2}$,
more compact results can be reexpressed as
\bea
\la x\ra_f = F(y)= \frac{(B+y)d}{1+By} \,,
\eea
while $\la x\ra_{\bar{f}} = F(-y)$.
Here we have introduced $B=\alpha^2-\beta^2$ and $y=\cos\theta$
to characterize the PPS states.
Similarly, the PSA and PSR probabilities are reexpressed as
$p_f=(1+By)/2$ and $p_{\bar{f}}=(1-By)/2$.

Following Refs.\ \cite{Bru13,ABWV16}, the DSA scheme considers
using the difference of the distribution functions,
i.e., $\widetilde{P}^{(-)}(x)= \widetilde{P}_1(x)-\widetilde{P}_2(x)$,
as a signal function from which the parameter is to be extracted.
Let us assume using $N$ particles in experiment.
The PSA and PSR distribution functions correspond to
\bea
\widetilde{P}_1(x) = \frac{n_1(x)}{N}
~~{\rm and}~~ \widetilde{P}_2(x) = \frac{n_2(x)}{N} \,,
\eea
where $n_1(x)$ and $n_2(x)$ are
the PSA and PSR particle numbers at point $x$.
Then, one can define the difference-signal as
\bea\label{diff-P-1}
P^{(-)}(x)=\frac{n_1(x)-n_2(x)}{N_1-N_2}   \,.
\eea
This is the normalized version of $\widetilde{P}^{(-)}(x)$,
with $N_1$ and $N_2$ the total PSA and PSR particle numbers.
Using $P^{(-)}(x)$, one can compute the average
$\bar{x} = \int dx\, x\, P^{(-)}(x)$, which gives
\bea\label{xbar-1}
\bar{x}
&=& \left(\frac{\delta_1}{\delta_1-\delta_2}\right) \la x\ra_{f}
- \left(\frac{\delta_2}{\delta_1-\delta_2}\right) \la x\ra_{\bar{f}}  \nl
&\equiv&  \beta_1  \la x\ra_{f} - \beta_2 \la x\ra_{\bar{f}}   \,,
\eea
where $\delta_1=N_1/N$ and $\delta_2=N_2/N$.
In experiment, the averages $\la x\ra_{f}$ and $\la x\ra_{\bar{f}}$
are determined using the distribution functions
$P_1(x)=n_1(x)/N_1$ and $P_2(x)=n_2(x)/N_2$;
while in theory, they are computed using \Eq{xf-xfbar}.
In theory, we also have $\beta_1=p_f/(p_f-p_{\bar{f}})$
and $\beta_2=p_{\bar{f}}/(p_f-p_{\bar{f}})$.
Simple calculation gives
\bea
\beta_1 &=& \frac{1+By}{2By} \,, \nl
\beta_2 &=& \frac{1-By}{2By}  \,.
\eea
Making contact between the experimental and theoretical results of $\bar{x}$,
one can extract (estimate) the value of the parameter $d$.
In classical case, the final theoretical result of $\bar{x}$ is
\bea\label{xbar-2}
\bar{x}=\beta_1 \, \la x\ra_f  - \beta_2 \, \la x\ra_{\bar{f}} = \frac{d}{B} \,.
\eea
This result is unexpectedly simple,
which is only determined by the pre-selection
but does not depend on the post-selection.
One can easily check that the simple reason of obtaining this result
is the cancellation of the post-selection factor $y$,
during multiplying $\beta_1 \, \la x\ra_f$ and $\beta_2 \, \la x\ra_{\bar{f}}$,
and making difference between them.
We may remark that in quantum case, this type of cancellation does not occur
and the resultant JWVA signal $\bar{x}$ depends on the post-selection.
In Fig.\ 1, we show the ratio factors $\beta_1$ and $\beta_2$,
and the averages $\la x\ra_f$ and $\la x\ra_{\bar{f}}$.
All of them depend on post-selection.

\begin{figure}
\includegraphics[scale=0.28]{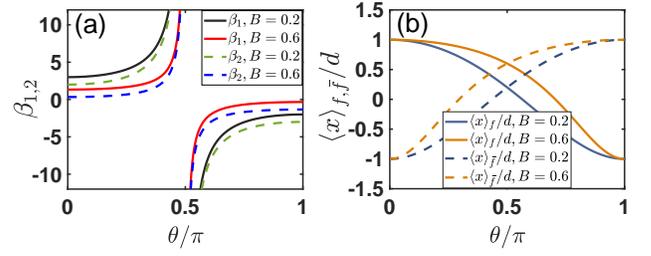}
\caption{Post-selection dependence of the ratio factors $\beta_{1,2}$ in (a)
and the individual averages $\la x\ra_{f,\bar{f}}$ in (b).
Here and in the following figures the pre- and post-selection states
are characterized by parameters $B$ and $\theta$
(see main text for their definitions).  }
\end{figure}

{\flushleft\bf Reformulation via difference-combined stochastic variables.}$~~$
Regarding the difference distribution function $P^{(-)}(x)$
as a probability function, we can compute the average $\bar{x}$.
However, we cannot use $P^{(-)}(x)$ to compute $\overline{x^2}$,
since this difference function is not {\it positive definite},
which may cause ill-behaved results,
e.g., making the statistical average of $x^2$ be negative.
This forbids us to know the uncertainty of $\bar{x}$
and thus to carry out the estimate precision in terms of signal-to-noise ratio.
To overcome this difficulty, let us consider each individual measured result $x_j$
as a specific ``realization" of the stochastic variable $\hat{x}_j$,
and group all the stochastic variables as follows
\bea
\hat{Y}_1 &=& \frac{1}{N_1} \sum^{N_1}_{j=1} \hat{x}^{(f)}_j  \,,  \nl
\hat{Y}_2 &=& \frac{1}{N_2} \sum^{N_2}_{k=1} \hat{x}^{(\bar{f})}_k  \,.
\eea
This corresponds to the experiment using $N$ particles,
with $N_1$ results accepted by the post-selection, and $N_2$ results rejected.
In the first group, each stochastic variable obeys the statistics
governed by $P_1(x)$ from \Eq{P1},
while in the second group each stochastic variable
obeys the statistics governed by $P_2(x)$ from \Eq{P2}.
Then, the {\it difference-signal} corresponds to the mean value
of the following difference-combined stochastic variables (DCSV) \cite{Li22}
\bea\label{SV-1}
\hat{x}&=&\left(\frac{N_1}{N_1-N_2}\right) \hat{Y}_1
- \left(\frac{N_2}{N_1-N_2}\right) \hat{Y}_2 \nl
&\equiv& ~ \beta_1 \hat{Y}_1 -\beta_2 \hat{Y}_2   \,.
\eea
The ensemble average of $\hat{x}$ reads as
\bea\label{SV-2}
{\rm E}[\hat{x}] = \beta_1 \la x\ra_f - \beta_2 \la x\ra_{\bar{f}} \,,
\eea
which is the same as the $\bar{x}$ calculated by using
the difference probability function $P^{(-)}(x)$.
However, the variance of $\hat{x}$
\bea\label{SV-3}
{\rm D}[\hat{x}] &=& \beta^2_1\, {\rm D}[\hat{Y}_1] + \beta^2_2\, {\rm D}[\hat{Y}_2]   \nl
&=& \beta^2_1 \left(\frac{\sigma^2_1}{N_1}\right)
    + \beta^2_2 \left(\frac{\sigma^2_2}{N_2}\right)   \,,
\eea
is now well-defined and {\it positive-definite},
which properly characterizes the estimate precision.
Here $\sigma^2_1$ and $\sigma^2_2$ are
the variances of the single stochastic variables
in the sub-ensembles defined by $P_1(x)$ and $P_2(x)$, respectively.
Simple calculation gives
\bea
\sigma^2_{1,2} = \sigma^2 + d^2
\left[ \frac{(1-B^2)\sin^2\theta}{(1 \pm B\cos\theta)^2} \right]  \,.
\eea
Here we have restored using the post-selection angle $\theta$ to express the result,
for the sake of being more compact.
We notice that, when $B\to 0$ and $\cos\theta\to 1$,
the variances of the sub-ensemble statistics coincide with
the original distribution width of the meter's wavepacket,
i.e., $\sigma^2_{1,2}\to \sigma^2$.
In general, the variances of the sub-ensemble results are shown in Fig.\ 2,
which depend on the pre- and post-selection choices,
as characterized by $B$ and $\theta$, respectively.

\begin{figure}
\includegraphics[scale=0.32]{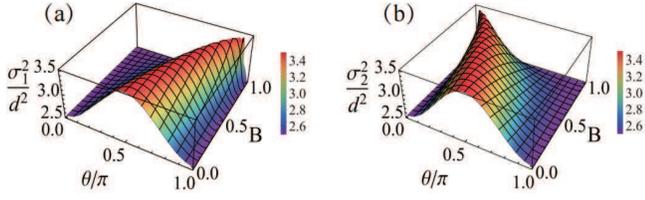}
\caption{
Variances of the stochastic variables
governed by the PSA and PSR sub-ensembles,
in (a) and (b), respectively.
Both are affected by the pre- and post-selection states.
The ratio value $\sigma^2/d^2=2.5$ is assumed in this plot.  }
\end{figure}

To characterize the quality of precision metrology,
following Refs.\ \cite{Ked12,Jor14,Li20},
we introduce the so called signal-to-noise ratio (SNR),
which is the ratio of the mean value $\bar{x}$
to the square root of the variance ${\rm D}[\hat{x}]$.
Here, the ``noise" corresponds to the uncertainty
of obtaining the mean value $\bar{x}$ (the ``signal").
Actually, this type of noise stems from the intrinsic shot noise,
while possible external technical noise is not accounted for in this work
but will be briefly discussed later near the end of this article.
We may remark that the SNR defined above is a reasonable figure-of-merit
to characterize the quality of the precision metrology,
viewing that the signal uncertainty alone
is not enough for the characterization,
since larger magnitude of the signal $\bar{x}$ is better for the metrological task.
Explicitly, we obtain the result of SNR as
\bea
&& R_{S/N} = \frac{|\bar{x}|}{\sqrt{{\rm D}[\hat{x}] }}  \nl
&& = 2\sqrt{N} |\cos\theta| \left[ \frac{g(1-B^2 \cos^2\theta)}
{4g(1-B^2)\sin^2\theta +(1-B^2\cos^2\theta)}  \right]^{1/2}  \nl
\eea
Here we introduced $g=(d/2\sigma)^2$,
which properly characterizes the measurement strength \cite{Ked12,Jor14,Li20}.

In Fig.\ 3, we show the numerical results of SNR
{\it versus} the PPS parameters $\theta$ and $B$ in (a),
and the feature of its weak dependence on the measurement strength $g$ in (b).
We notice that the overall behavior of the SNR plotted here
is quite similar to that of the quantum JWVA \cite{Li22}.
For instance, the SNR approaches to zero at $\theta=\pi/2$.
And, the quantum JWVA also has the feature of
weak dependence on the measurement strength $g$ \cite{Li22},
while the standard quantum WVA is quite sensitive to the strength \cite{Li20}.
The numerical results are scaled by the SNR of conventional measurement
(the optimal result without post-selection \cite{Ked12,Jor14,Li20,Li22a}),
i.e., $R_{S/N}^{\rm (cm)}=\sqrt{N} d/\sigma$.
Notice that for arbitrary initial state (arbitrary $B$),
when $\cos\theta\to \pm 1$ (i.e. $\theta=0$ or $\pi$),
the SNR of the DSA scheme approaches to $R_{S/N}^{\rm (cm)}$,
which is the upper bound achievable,
valid also for the quantum WVA and JWVA schemes in most cases \cite{Ked12,Jor14,Li20,Li22}
--- so far we only notice the exception when employing the optical
coherent state as a meter, which makes the SNR of the quantum WVA
possible to exceed the conventional scheme \cite{Li22a}.
The result in Fig.\ 3(a) is of great interest:
the SNR does not depend on $B$,
while the signal amplification is only determined by $B$.
This feature provides an important data-processing scheme in practice.
That is, we are allowed to choose the post-selection at $\theta=0$ or $\pi$
(in classical case which are equivalent)
and make the pre-selection parameter $B$ small.
Then, we can realize a large amplification for the signal
while at the same time keep the optimal SNR,
just as the WVA at the AAV limit \cite{Ked12,Jor14,Li20}.

\begin{figure}
\includegraphics[scale=0.37]{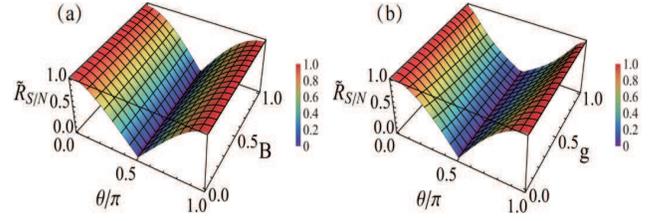}
\caption{
Numerical results of the SNR {\it versus} the pre- and post-selection
parameters $B$ and $\theta$ in (a),
and the feature of weak dependence on the measurement strength $g$ in (b).
The reduced results $\widetilde{R}_{S/N}=R_{S/N} / R_{S/N}^{\rm (cm)}$
are plotted by scaling with the SNR of conventional measurement
$R_{S/N}^{\rm (cm)}=\sqrt{N} d/\sigma$ (the optimal result without post-selection).
In (a) and (b), $g=0.1$ and $B=0.2$ are assumed, respectively.   }
\end{figure}

{\flushleft\bf Biased DSA scheme.}$~~$
Roughly speaking, for either the quantum case
in Refs.\ \cite{Bru13,ABWV16,ABWV17a,ABWV17b,Zeng16,Zeng19,Li22}
or the classical case analyzed in present work,
the principle of anomalous amplification is making the difference
of the PSA and PSR particle numbers
approach to zero, say, $N_1-N_2\to 0$.
This fact raises an interesting question: if the PPS design leads to $N_1 \neq N_2$,
can we handle the PSA and PSR results better in order to achieve a larger amplification?
Intuitively, based on the fact that $N_1-\eta N_2\to 0$,
where $\eta=p_f/p_{\bar{f}}$, we may consider the following
{\it biased} DSA (BDSA) scheme by constructing
\bea\label{diff-P-2}
P_{\beta}^{(-)}(x)=\frac{n_1(x)-\beta\, n_2(x)}{N_1-\beta N_2}   \,.
\eea
Then, large amplification seems possible when setting $\beta\to\eta$.
Similar to the unbiased DSA scheme discussed above,
we can construct also the DCSV as
\bea
\hat{x}_{\beta}&=&\left(\frac{N_1}{N_1-\beta N_2}\right) Y_1
- \left(\frac{\beta N_2}{N_1-\beta N_2}\right) Y_2 \nl
&\equiv& ~ \tilde{\beta}_1 Y_1 -\tilde{\beta}_2 Y_2   \,.
\eea
Based on this biased DCSV formulation, we can straightforwardly
carry out the mean value and variance as follows.
The mean value is
\bea\label{xbar-beta}
\bar{x}_{\beta} &=& \frac{1}{\eta-\beta} (\eta\la x\ra_f - \beta\la x\ra_{\bar{f}})  \nl
&\simeq& \frac{\eta}{\eta-\beta} [F(y)-F(-y)]  \,.
\eea
The result of the second line is from considering
the amplification condition $\beta\simeq \eta$.
The variance is obtained as
\bea
{\rm D}[\hat{x}_{\beta}]
&=& \tilde{\beta}^2_1 \left(\frac{\sigma^2_1}{N_1}\right)
    + \tilde{\beta}^2_2 \left(\frac{\sigma^2_2}{N_2}\right)    \nl
&=& \frac{1}{N} \left( \frac{p_f}{p^2_{\beta}}\sigma^2_1
+ \beta^2\frac{p_{\bar{f}}}{p^2_{\beta}}\sigma^2_2   \right)   \nl
&\simeq & \frac{1}{N} \frac{p_f}{p^2_{\beta}} (\sigma^2_1 + \eta \sigma^2_2) \,.
\eea
Here we have introduced $p_{\beta}=p_f-\beta p_{\bar{f}}$
and considered also the condition $\beta\simeq \eta$.
Then, the SNR of the BDSA is given by
\bea\label{R-SN-b}
R^{\rm (b)}_{S/N} = \frac{\bar{x}_{\beta}}{\sqrt{D[\hat{x}_{\beta}]}}
\simeq \sqrt{N p_f}
\left(\frac{\widetilde{F}(y) }{\sqrt{\sigma^2_1 + \eta\sigma^2_2}} \right) \,,
\eea
while $\widetilde{F}(y)$ is defined as
\bea\label{dFy}
\widetilde{F}(y)\equiv F(y)-F(-y)=\frac{2y(1-B^2)}{1-B^2y^2} d  \,.
\eea
In Fig.\ 4 we show the effect of signal amplification by the BDSA technique.
Unlike the unbiased DSA scheme, as shown by the simple result of \Eq{xbar-2},
here the signal amplification is no longer independent of the post-selection.
In particular, the singular amplification corresponds to
the condition $\eta \simeq \beta$.
Thus we understand that for different $B$ (different pre-selected state)
the singular amplification occurs at different $\theta$
(different post-selection angle), as shown in Fig.\ 4.
However, the signal's amplification is accompanied
with enhancement of uncertainty of the signal.
In Fig.\ 5, we show the SNR of the BDSA,
under different choice of the bias parameter $\beta$.
We find that the overall behavior of SNR for different $\beta$
is similar to each other.
Remarkably, for this BDSA strategy,
the singular amplification shown in Fig.\ 4 does not indicate
that we can get the optimal (maximum) SNR under this condition.
The reason is that when we consider the SNR,
the amplification factor $\frac{1}{\eta-\beta}$
would be canceled from the numerator and denominator in the ratio.
Then, the resultant value of the SNR is largely determined
by the factor $\widetilde{F}(y)$, as shown by \Eqs{R-SN-b} and (\ref{dFy}).
Obviously, the condition of singular amplification
does not coincide with the condition of maximum $\widetilde{F}(y)$.
Qualitatively speaking, it is also this factor
that results in the {\it zero-lines} of the SNR shown in Fig.\ 5,
despite that, quantitatively,
the zero-lines weakly depend on the bias parameter $\beta$.
We notice that the maximum SNR is obtained also at $\theta=0$ and $\pi$,
while the precise value of SNR is slightly smaller than $R^{(\rm cm)}_{S/N}$
and is weakly affected by $B$ and $\beta$.
However, this BDSA cannot realize large amplification
by reducing $B$ at $\theta=0$ or $\pi$ (as shown in Fig.\ 4).
Connecting the results shown in Figs.\ 4 and 5, from the perspective of
getting large amplification {\it and} keeping optimal SNR,
we may conclude that the simple unbiased DSA scheme should be better than the BDSA scheme,
despite that the latter can realize large amplification of signal
aside from the limit $B\to 0$, as shown in Fig.\ 4.

\begin{figure}
\includegraphics[scale=0.45]{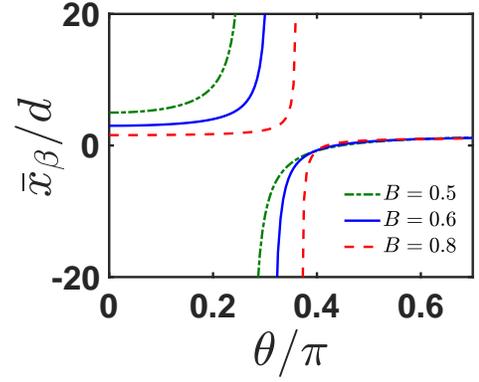}
\caption{
Singular amplification behavior in the BDSA scheme,
by matching the ratio $\eta=p_f/p_{\bar{f}}$
of the PSA and PSR probabilities
with the bias parameter $\beta$ ($\beta=2$ in this plot). }
\end{figure}

\begin{figure}
\includegraphics[scale=0.32]{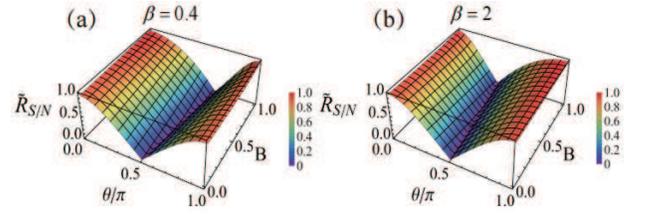}
\caption{
SNR of the BDSA (reduced results as in Fig.\ 3),
for bias parameters $\beta=0.4$ and 2 in (a) and (b).
The measurement strength $g=0.1$ is assumed in this plot.   }
\end{figure}

{\flushleft\bf Discussion.}$~~$
In quantum case, the post-selection with a quantum pure state
can be quite naturally implemented by a quantum projective measurement
on such as the spin of an electron or polarization of a photon.
However, post-selection with a mixed state will be more tricky.
As proposed in Ref.\ \cite{Vaid17},
a possible way is to couple the system of interest
(e.g. a two-state system with states $|1\ra$ and $|2\ra$) to an ancilla,
and to prepare them in an entangled state
$|\Psi_T\ra=a|1\ra |\chi_1\ra + b|2\ra |\chi_2\ra$.
If we keep the ancilla being protected to avoid any specific observation/measurement,
the resultant system state is $\rho_s={\rm Tr}_{\rm an}(|\Psi_T\ra \la\Psi_T|)$,
where ${\rm Tr}_{\rm an}(\cdots)$ means averaging the ancilla state
(i.e., ``ignoring" the ancilla).
If the two states of the ancilla are orthogonal to each other,
i.e., $\la \chi_1|\chi_2 \ra=0$,
we then obtain a fully classical mixed state as
$\rho_s=a^2|1\ra\la 1| + b^2|2\ra\la 2|$.

Actually, the post-selection can be handled as a {\it post-processing of data}.
Holding the recorded distribution $n(x)$ of the output results in experiment
and guided by the theoretical probabilities
$\widetilde{P}(x;f)$ and $\widetilde{P}(x;\bar{f})$,
we can simply obtain the PSA and PSR distributions $n_1(x)$ and $n_2(x)$
by computing
$N \widetilde{P}(x;f)=n_1(x)$ and $N \widetilde{P}(x;f)=n_2(x)$.
Then we have $P_1(x)=n_1(x)/N_1$ and $P_2(x)=n_2(x)/N_2$ and use them
to compute the averages $\la x\ra_f$ and $\la x\ra_{\bar{f}}$.
From \Eq{xbar-1}, we obtain $\bar{x}$
and can estimate the parameter $d$ from this amplified signal
by using the simple relation $\bar{x}=d/B$ of \Eq{xbar-2} .
After knowing the averages $\la x\ra_f$ and $\la x\ra_{\bar{f}}$
from the experimental data,
one can also utilize the BDSA signal $\bar{x}_{\beta}$
to extract the parameter $d$, based on \Eq{xbar-beta}.

We may highlight some technical advantages of the DSA as follows.
{\it (i)}
The most prominent advantage of the DSA should be the possibility of removing
some systematic errors such as misalignment imperfection.
Let us imagine that the PSA signal $n_1(x)$ and the PSR signal $n_2(x)$
shift towards the same direction, owing to an error of misalignment.
The subtracting procedure in the DSA
would eliminate this error from the difference signal $n_1(x)-n_2(x)$.
Thus the common shift of $\la x\ra_f$ and $\la x\ra_{\bar{f}}$ will be eliminated.
{\it (ii)}
In the quantum WVA, only a small portion of output results are remained,
thus the signal is very weak.
However, the flux intensity of particles cannot be so weak
owing to the limitation from some imperfections in the post-selection process.
By accounting for this intensity limitation, it was estimated in Ref.\ \cite{Zeng19}
that the JWVA scheme can outperform the standard WVA approach
by two orders of magnitude higher in sensitivity.
Since the DSA analyzed in present work does not differ too much from the quantum JWVA,
we thus expect the classical DSA technique
to share the same advantage of the JWVA as pointed out in Ref.\ \cite{Zeng19}.
{\it (iii)}
The WVA, JWVA, and DSA can outperform conventional scheme
(without post-processing of the output data) beyond detector's resolution.
Obviously, if the signal shift falls into the range of detector's resolution,
conventional scheme will fail.
However, as already demonstrated in the WVA
\cite{Kwi08,How09a,How09b,How10a,How10b,Guo13}
and JWVA \cite{ABWV16,ABWV17a,ABWV17b,Zeng16,Zeng19} experiments,
tiny shifts beyond detector's resolution can be measured.
We expect the DSA technique to hold similar ability,
even for precision metrology in classical systems,
since the amplified signal $\bar{x}=d/B$
can drastically exceed detector's resolution as well.

To summarize, in this work we have proposed and analyzed a quantum-coherence-free
amplification scheme of precision metrology, termed as DSA.
Our analysis was based on the Stern-Gerlach setup
by erasing quantum coherence of the electron's spin.
We obtained a simple expression for the amplified signal,
carried out characterization of estimate precision,
and pointed out the optimal working regime.
We also discussed how to implement the post-selection of a classical mixed state.
The proposed DSA scheme may find valuable applications in practice.

\vspace{0.8cm}
{\flushleft\it Acknowledgements.}---
This work was supported by the NNSF of China
under Nos.\ 11974011 and 61905174.


\end{document}